\newcommand{\DocVersion}{Accepted for publication in ApJL, 4 Nov 2013.}

\documentclass[preprint,11pt]{aastex}

\slugcomment{\DocVersion}

\shortauthors{Shuping et al.}
\shorttitle{Orientation and morphology of HST-10}

\newcommand{\kms}{km~s$^{-1}$}

\newcommand{\cmthree}{cm$^{-3}$}

\newcommand{\msun}{M$_{\sun}$}

\newcommand{\htwo}{H$_2$}

\newcommand{\brgamma}{Br~$\gamma$}
\newcommand{\thetaonec}{$\theta^1$~C}
\newcommand{\thetatwoa}{$\theta^2$~A}
\newcommand{\halpha}{H$\alpha$}
\newcommand{\helium}{\ion{He}{1}}
\newcommand{\Oone}{\ion{O}{1}}

\begin{document}

\title{The curious morphology and orientation of Orion proplyd HST-10}

\author{R. Y. Shuping\altaffilmark{1}}
\affil{Space Science Institute \\
4750 Walnut St., Suite 205, \\
Boulder, CO   80301}
\email{rshuping@spacescience.org}

\and

\author{Marc Kassis}
\affil{W.M. Keck Observatory \\
65-1120 Mamalahoa Hwy. \\
Kamuela, HI    96743-8431
}

\and

\author{John Bally}
\affil{Center for Astrophysics \& Space Astronomy \\
Univ. of Colorado \\
391 UCB \\
Boulder, CO 80309-0001
}

\and

\author{Mark R. Morris}
\affil{Department of Physics, Astronomy Division \\
University of California, Los Angeles \\
PO Box 951547 \\
Los Angeles, CA 90095-1547}

\altaffiltext{1}{USRA-SOFIA, NASA-Ames Research Center, MS N232-12, PO Box 1, Moffett Field, CA 94035}

\begin{abstract}

HST-10 is one of the largest proplyds in the Orion Nebula and is located approximately 1\arcmin\ SE of the Trapezium.  Unlike other proplyds in Orion, however, the long-axis of HST-10 does {\em not} align with \thetaonec, but is instead aligned with the rotational axis of the HST-10 disk.  This cannot be easily explained using current photo-evaporation models.  In this letter, we present high spatial resolution near-infrared images of the Orion proplyd HST-10 using Keck/NIRC2 with the Laser Guide Star Adaptive Optics system, along with multi-epoch analysis of HH objects near HST-10 using Hubble Space Telescope WFPC2 and ACS cameras.  Our narrow-band near-IR images resolve the proplyd ionization front (IF) and circumstellar disk down to 23~AU at the distance to Orion in \brgamma, \helium, \htwo, and PAH emission.   \brgamma\ and \helium\ emission primarily trace the IF (with the disk showing prominently in silhouette), while the \htwo\ and PAH emission trace the surface of the disk itself.  PAH emission also traces small dust grains within the proplyd envelope which is asymmetric and does not coincide with the IF.  The curious morphology of the PAH emission may be due to  UV-heating by both \thetaonec ~Ori and \thetatwoa ~Ori.  Multi-epoch HST images of the HST-10 field show proper motion of 3 knots associated with HH~517, clearly indicating that HST-10 has a jet.  We postulate that the orientation of HST-10 is determined by the combined ram-pressure of  this jet  and the FUV-powered photo-ablation flow 
from the disk surface.

\end{abstract}

\keywords{stars: formation---stars: jets---stars: winds, outflows---circumstellar matter}


\section{Introduction}
\label{Intro}

Disks surrounding young stellar objects (YSOs) are critical to our understanding of star formation.  Not only do circumstellar disks provide a mechanism for angular momentum transport and accretion onto the central star during formation, but they also serve as the birthplace of planets.  The {\it Hubble Space Telescope} has produced dramatic images of protoplanetary disks (``proplyds'') surrounding young ($< 10^6$~year old) stars in the Orion Nebula~\citep{Odell:1993,Bally+98,Bally:2000}.  The intense ultraviolet (UV) radiation field of the high-mass Trapezium stars heats the disk surfaces to a few thousand degrees, which drives a photo-ablation flow, and produces bright ionization fronts~\citep{Johnstone:1998}. 

Disk mass-loss rates in YSOs due to external photoevaporation place strong temporal constraints on planet formation mechanisms in irradiated environments near O-type stars; which is important since a significant fraction of stars forms in clusters with high-mass stars~\citep{Adams:2010}.  Furthermore, mounting evidence---from solar system size to meteoritics---now strongly suggests that our own Solar System in fact formed in a large stellar cluster, perhaps similar to the ONC~\citep[and references therein]{Gounelle:2012,Gritschneder:2012,Adams:2010a,Young:2011}. So it is clearly important to gain an understanding of how efficient planet formation is for low-mass stars forming in clusters with high-mass members, both to predict the frequency of planetary systems and to better understand the evolution of our own solar system.  

The formation and evolution of dust grains in proplyd systems is  key to understanding overall system evolution and whether or not the disks might produce planets.  Like other circumstellar disks around young stars, it is thought that dust grains will grow via accretion and agglomeration.  Dust grains are also thought to be entrained in the photoevaporative flow off the disk, effectively removing material from the grain-growth process.  If the photoevaporative entrainment is very efficient, then we would not expect to see evidence for larger grain sizes.  Mid-IR observations of the Orion proplyds closest to \thetaonec\ clearly show strong emission both from the proplyd itself, and from the large wind-wind shock fronts where the photoevaporation flow rams into the stellar wind from \thetaonec\ \citep{Smith:2005,Robberto:2005}, suggesting that the grains are indeed entrained in the photoevaporative flow.  At the same time, in the case of proplyd \object[ow94 114-426]{114-426}, measurements of the extinction curve through the outer parts of the disk indicate grain growth to sizes larger than typical in the ISM~\citep{Throop+01,Shuping:2003, Miotello:2012}.  Furthermore, silicate emission profiles observed for a number of proplyds are similar to other low-mass YSOs and reveal that the grains are both larger than normal ISM sizes  and chemically processed~\citep{Shuping:2006}.   Though it appears that dust is entrained in the photoevaporative flow, it is still not clear whether this has a significant effect on the growth of grains in the circumstellar disk.  

\object[ow94 182-413]{HST-10 (182-413)} is one of the largest proplyds in Orion:  the outer cometary-shaped ionization front (IF) is roughly 1\farcs 0 $\times$ 2.5\arcsec\ in size---corresponding to approximately 420~AU $\times$ 1000~AU at the distance of Orion\footnote{
We adopt a distance of 416$\pm6$~pc~\citep{Kim:2008} throughout the paper.}---making HST-10 one of the few proplyds that can be resolved from the ground.  Unlike the proplyds near \thetaonec , HST-10 has no bright stand-off shock due to wind-wind collision.  At the center of the cometary envelope is an edge-on silhouette disk with a radius of 0.18\arcsec ~\citep{Bally+98}, corresponding to a radius of 75~AU, with a disk mass of $(7.4\pm0.7) \times 10^{-3}$~\msun\ ~\citep{Mann:2009}. Though dark in continuum and \halpha\ images, the disk glows in both \Oone\ and \htwo\ \citep{Chen+98,Bally:2000}, indicating that the photoevaporative flow off the disk is driven by the far-UV rather than the extreme-UV radiation field~\citep{Johnstone:1998}.  One of the most notable features of HST-10 is that---unlike other proplyds in Orion---the long-axis of the envelope does {\em not} point toward \thetaonec\ but instead appears to be aligned with the rotation axis of the disk.  

In this letter, we present high spatial resolution ground-based near-infrared images of the Orion proplyd HST-10 using Keck/NIRC2 with the Laser Guide Star Adaptive Optics system that resolves the IF and disks down to size scales of 23~AU.  We also present multi-epoch analysis of HH objects near HST-10 that are aligned with the rotational axis of the silhouette disk---clearly indicating that HST-10 has a strong outflow component.  These observations indicate that the orientation and morphology of HST-10 is somewhat more complex than previously thought.

\section{Near-IR Narrow Band Imaging}

\begin{figure}[e]
   \begin{center}
   \epsscale{0.33}
   \plotone{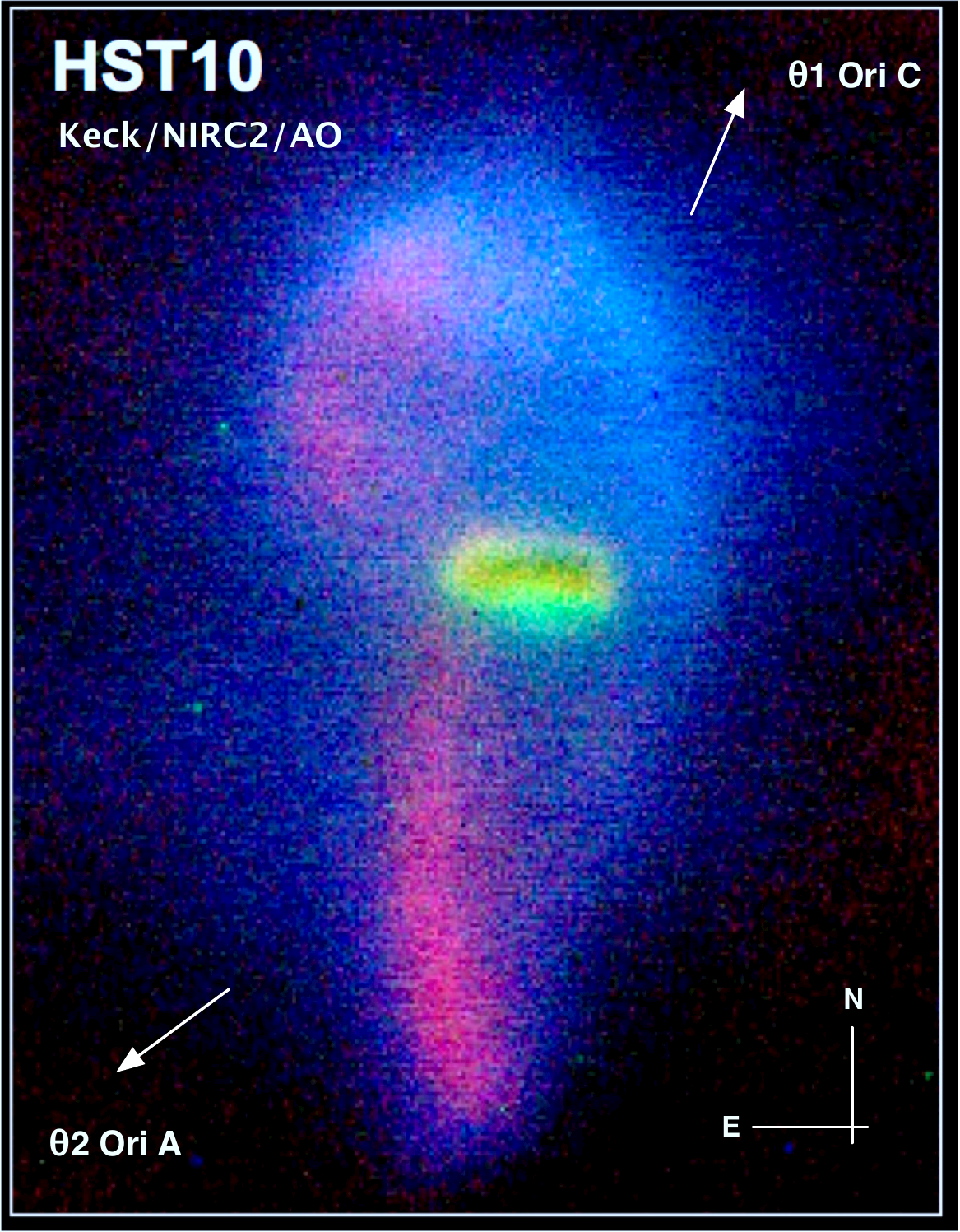}
    \epsscale{0.5}
   \plotone{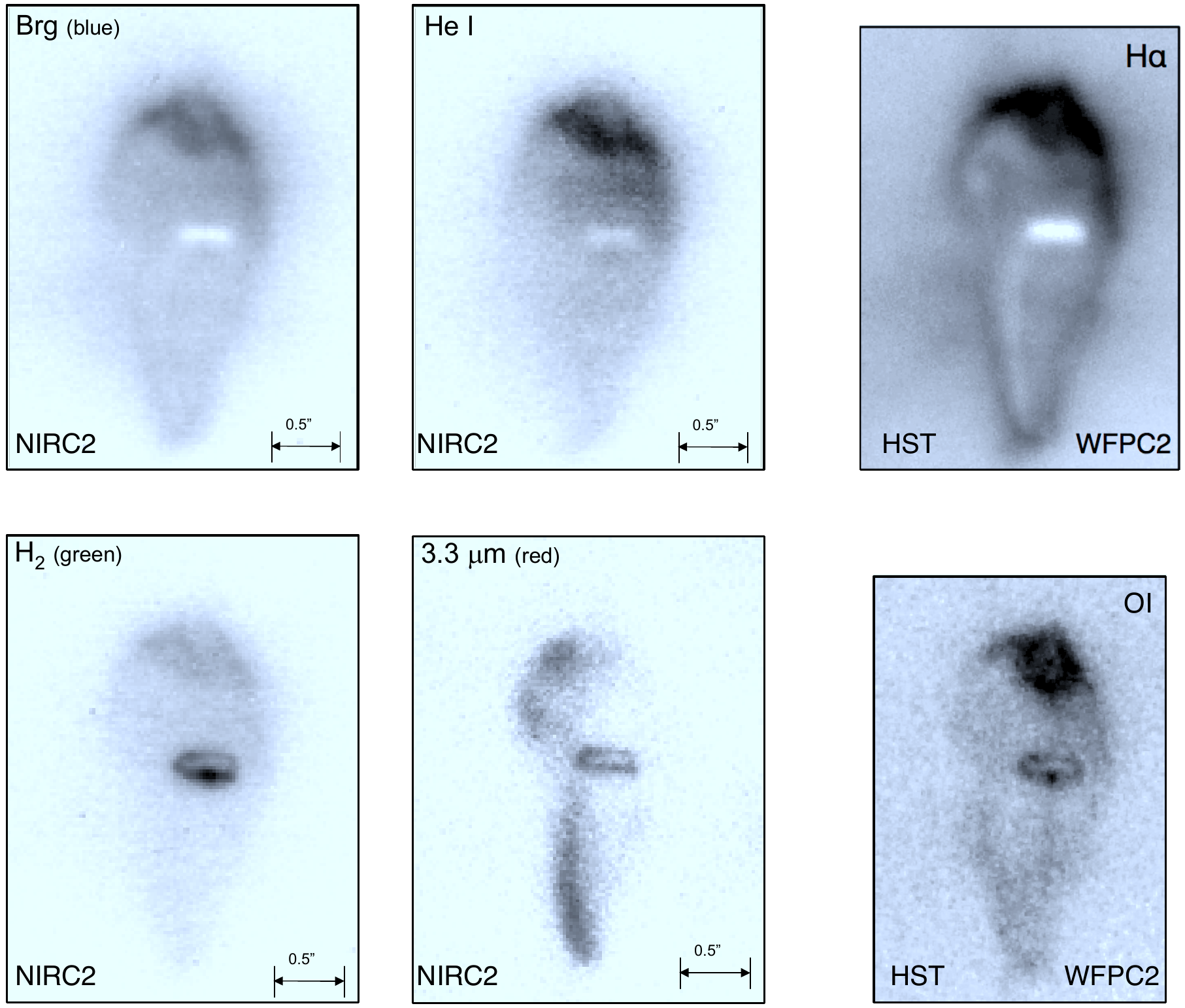}
   \end{center}
   \caption[Proplyd HST-10 (NIRC2/LGS-AO)] {Keck/NIRC2 LGSAO imaging of HST-10 in the Orion Nebula. Grey-scale images are displayed using a logarithmic intensity stretch for all observed wavelengths: \brgamma\ (2.17~\micron), \ion{He}{1} (2.06~\micron), H$_2$ (2.12~\micron), and PAH (3.3~\micron). Each image represents 15 min. of on-source time.  \halpha\ and OI narrow-band HST/WFPC2 images~\citep[Cycle~6,][]{Bally:2000} are shown on the right-hand side for comparison.
\label{fig:HST10_nirc2} 
}
   \end{figure}

Narrow-band imaging observations of HST-10 were acquired with NIRC2-Keck~II in LGS-AO mode on 17 Nov 2006 using Keck Observatory director's discretionary time.  Data were acquired in the NIRC2 narrow camera providing a scale of 10 mas/pixel (FOV: 10\arcsec). Observations were carried out using a 3-point dither pattern with a total integration time of 15~min for each filter.  Natural seeing at the beginning of the night was 0.85\arcsec\ in K-band. FWHM of the AO-corrected PSF was 80~mas (34~AU at Orion) with a 10-15\% Strehl ratio. Spatial resolution in our images of HST-10 are the among the highest available from the ground and are comparable to optical images achieved by HST.  Flat-fielding, sky subtraction, and bad-pixel replacement were handled using standard near-IR techniques.    The final images for each filter are shown in Figure~\ref{fig:HST10_nirc2}.

Both \brgamma\ and \helium\ emission are brightest in the northern ``head'' of the HST-10 envelope and exhibit the same morphology as seen in HST images of similar spatial resolution~\citep{Bally:2000}.  Emission from the ionized gas appears to peak on the side of the envelope facing \thetaonec , the dominant source of EUV radiation.  The HST-10 envelope shows a slight limb brightening in \brgamma\ that is not apparent in the \helium\ emission, as expected for an ``inside-out'' PDR structure.  The 0\farcs 4 diameter circular structure apparent at the head of the proplyd is also apparent in \halpha\ and especially O~[I] HST images~\citep{Bally:2000}.  This circular structure may be due to the interaction between a jet (also apparent in O~[I]) and the edge of the proplyd envelope~\citep[e.g.][]{Bally:2005a}.

The HST-10 proto-planetary disk is resolved at all observed wavelengths (Fig.~\ref{fig:HST10_nirc2}). While the edge-on disk remains dark in images of the ionized gas, it is bright in the H$_2$ and PAH (3.3~\micron) filters.  Furthermore, the \htwo\ and PAH emission appear to come from a thin ``skin'' on the surface of the disk which is spatially resolved at these wavelengths for the first time.  The intensity of both the H$_2$ and PAH emission is clearly brighter on the south and east edges of the disk, which could simply be an inclination effect.  However, the discrepancy may also be due to differences in physical conditions at the surface of the disk, which will be discussed further below.  \htwo\ and \Oone\ emission from the disk are nearly identical in morphology, including a significant ``knot'' of emission in the center of the southern edge thought to be associated with shock emission from a neutral jet~\citep{Bally:2000}.

The PAH emission from the disk is similar to that seen in \Oone\ and \htwo\ except that no knot is seen on the south side, presumably because PAHs cannot survive in the shock-excited region responsible for the knot.   The morphology of the PAH emission beyond the disk differs significantly from that of the extended  gas and  dust.  The PAH emission at 3.3~\micron\ appears to trace the inner edge of the IF on the East side of the central disk and is nearly identical to the morphology observed in the 11.25~\micron\ PAH band~\citep[PAH2, ][]{Vicente:2013}.    As noted by \citet{Vicente:2013}, the PAH emission is also well-correlated with the dark features seen in \halpha\ images (Fig.~\ref{fig:HST10_nirc2}) which  suggests that these dark lanes are due primarily to small grains.   The overall asymmetry of the PAH emission in the HST-10 envelope will be discussed below.  

\section{Dust in the HST10 Photoevaporative Flow}

Observed proplyd morphology is a consequence of photoevaporation of the circumstellar disk by UV radiation from one  side (\thetaonec\ in this case) in addition to isotropic nebular UV radiation.  The combination of these effects produces the familiar IF head-tail morphology aligned with \thetaonec .  The photoevaporative flow from a disk subjected to {\em both} strong FUV and EUV radiation can be modeled as two coupled Parker winds:  a supersonic {\em neutral} wind off the disk and a second supersonic {\em ionized} wind off the IF~\citep[and references therein]{Clarke:2011}. If the IF is {\em beyond} the sonic point for the neutral wind (as is the case for HST-10), then a ``shocked neutral shell'' is formed just inside of the IF where the gas is down-shocked to subsonic velocities before passing through the IF.  Dust grains can be entrained in the photoevaporative flow only if the drag force on the grain (due to the gas) is greater than the gravitational force, {\em and} the grain has sufficient time to reach escape velocity from the disk~\citep[see][]{Adams:2004}.  Hence, entrainment is much more likely for small grains.  PAH emission from within the proplyd envelope of HST-10 indicates that small grains are indeed entrained in the flow off the disk~\citep[this work]{Vicente:2013}.

The morphology of the PAH emission in HST-10, however, is surprising.  If we adopt the model above, we would expect a symmetric distribution of PAH emission tracing the neutral shell just inside the IF---but this is not observed.  Instead, the PAH emission is just inside the IF on the {\em east} side of the envelope, and non-existent on the west side (this is also evident in the dark extinction lanes visible in the \halpha\ HST image, see Fig.~\ref{fig:HST10_nirc2}).  The asymmetry is unlikely to be caused by PAH ionization since the 8.6~\micron\ ionized PAH emission reported by \citet{Vicente:2013} also appears primarily on east side of the envelope.  

If we assume that PAHs are indeed entrained in the flow from the northern face (and edges) of the disk, then the lack of PAH emission from the NW quadrant of the proplyd ``head'' (toward \thetaonec ) could be due to the ``hard'' FUV radiation that can penetrate the IF into the neutral PDR, causing either PAH destruction (via 2-photon UV absorption) or dissociation of the outer C-H bond that is responsible for 3.3~\micron\ emission.  The PAH emission in the southern ``tail'', however, is more problematic:  if we assume that the flow off the south side of the disk is driven by the isotropic nebular UV radiation, then it is hard to explain why the PAH emission is only apparent on the {\em east} side of the envelope.  

An alternative explanation is that  HST-10 is subjected to an additional source of FUV radiation from the nearby early-B star \thetatwoa\ (82\arcsec\ to the SE) which drives a photoevaporative flow off both the southern face and eastern edge that is ionized by the diffuse nebular radiation a few disk radii away.  PAHs are entrained in the flow off the east edge---creating a ``tail'' of PAH emission in the SE quadrant of the neutral envelope---but are {\em not} entrained in the flow off the south face of the disk, perhaps because gas drag forces cannot overcome gravity.   Hence, there is little or no PAH emission from the SW quadrant of the neutral envelope.   Clearly, additional observations are needed to better understand PAH emission in photo-evaporating disk systems, but it seems plausible that additional sources of UV radiation may need to be invoked when modeling photo-ablation flows in proplyd systems.

\section{Proper Motions of HH~517}

Using observations obtained with HST WFPC2 in 1998, \citet{Bally:2000} reported the 
presence of a 6300 \AA\ \ion{O}{1}  jet emerging orthogonal to the 
HST-10 disk on its south  side that was designated  HH~517, as well as a pair of compact  H$\alpha$ knots located roughly 7\arcsec\ and 10\arcsec\ due  South.  Knot s1 has a faint tail extending back towards HST-10.  In addition to these features, the northern side of HST-10
exhibits a ring of emission in the IF along the expected jet axis.  As shown in
Figure~\ref{fig:HST10_nirc2} , faint spikes of H$\alpha$ emission extend above the 
proplyd IF at the eastern and western ends of this ring:
These features may mark the walls of an outflow cavity protruding beyond
the HST-10 IF. 

\begin{figure}[e]
   \begin{center}
   \epsscale{1.0}
   \plottwo{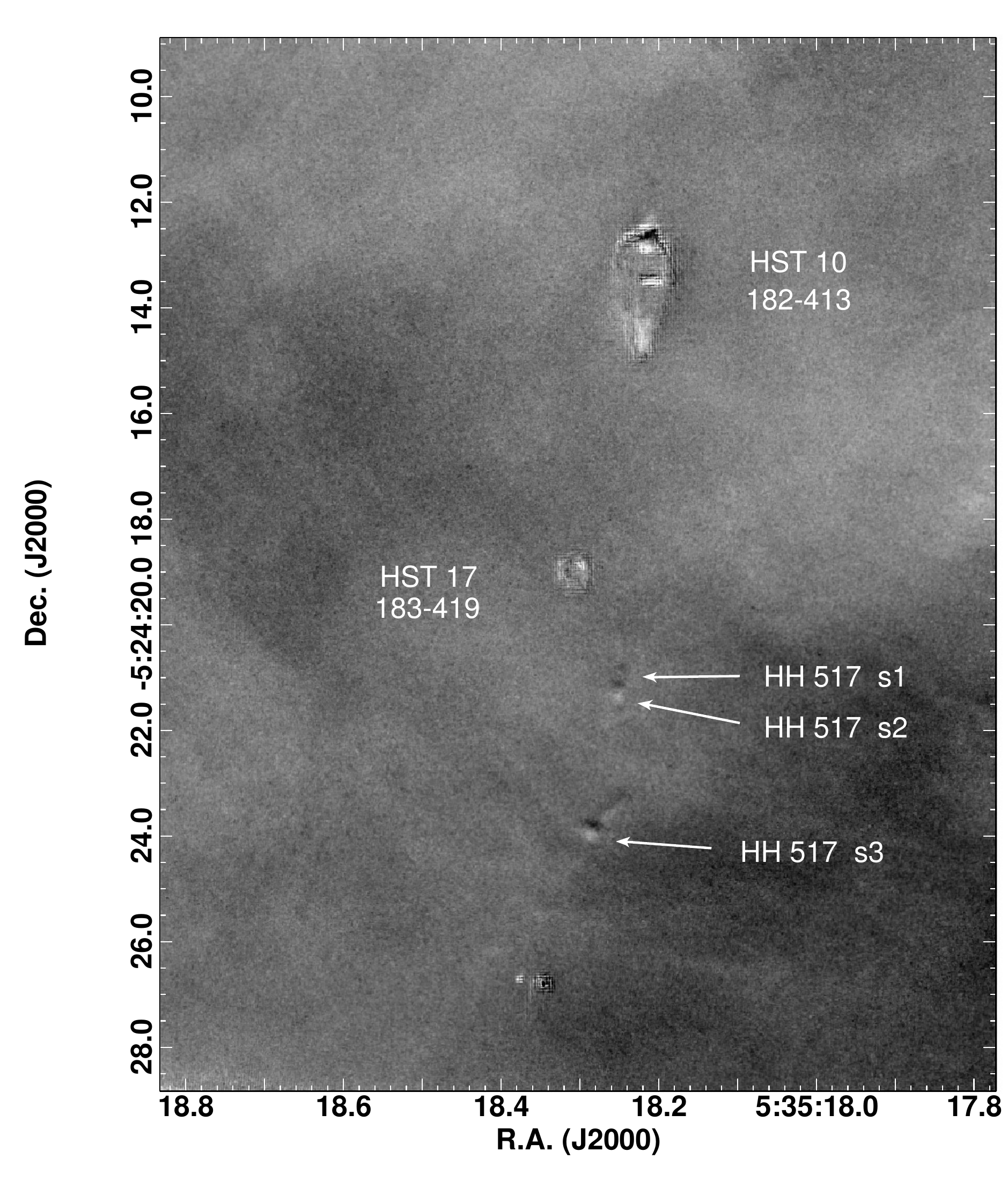}{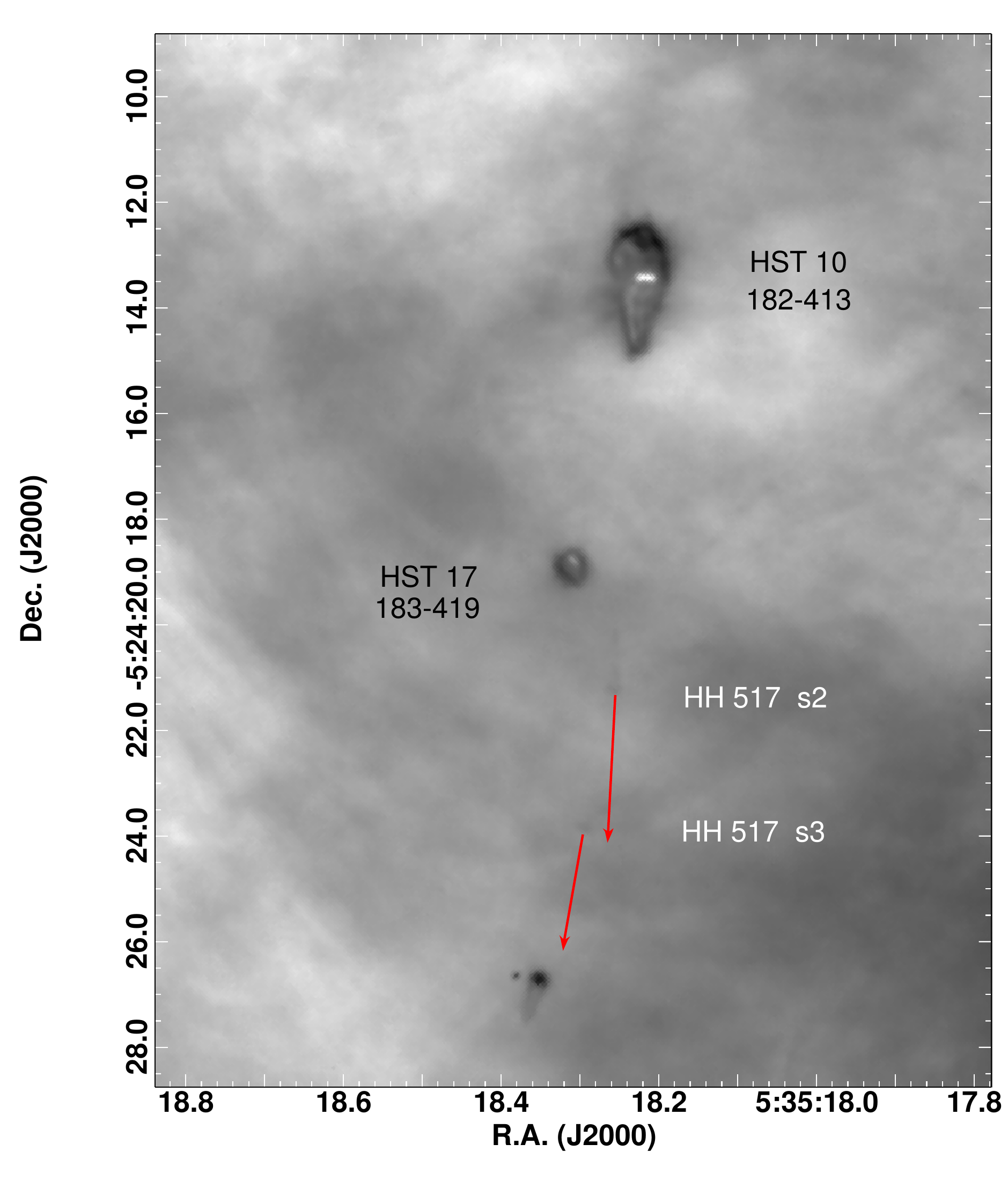}
   \end{center}
   \caption[HH~517 Proper Motions] {Left---Intensity matched, difference of registered  ACS \halpha\ images taken in 2005 (Cycle~13) and 1998 (Cycle~6).  The observed proper motions of knots HH~517 s1, s2, and s3 are clearly indicated
by the black-white ``dipoles''.  Right---HST/ACS H$\alpha$ image of HST-10 region obtained in Cycle~12.  Arrows are
 inferred proper motion vectors showing expected distance traveled in a time interval of 100~years.  
\label{fig:propermotions}}
   \end{figure}

Subsequent HST observations in Cycles~12 and 13 using the ACS/WFC  (see Table~\ref{table:hstobservations}) show that the  knots have large proper motions, indicating movement directly away from HST-10.   
Observed proper motion of the knots is clearly seen in a difference image between Cycles~13 and 6 (Fig.~\ref{fig:propermotions}).  Table~\ref{table:propermotions} gives the J2000
coordinates of the photo-centers of the knots as measured
on the 2004 ACS H$\alpha$ images along with the measured proper
motions based on the three epochs of observations. Inferred proper motion vectors are shown in Figure~\ref{fig:propermotions}.  The uncertainties in the velocities and vector orientations are  $\sim 10$\%.

\begin{deluxetable}{cccccc}
\tabletypesize{\footnotesize}
\tablewidth{0pt}
\tablecaption{HST Observations of HST-10\label{table:hstobservations}}
\tablehead{
\colhead{Cycle}  & \colhead{UT Date} & \colhead{Julian Date}& \colhead{Instrument}  &  \colhead{Filters} &
 \colhead{Program/PI} 
}
\startdata					
6  &  11~Feb~1998  &  51120  &  WFPC2  &  F631N, F656N  &  GO~6603/J.~Bally  \\
12 & 24~Jan~2004  &  53028  &  ACS/WFC  &  F658N  &  GO~9825/J.~Bally  \\
13 & 8~Apr~2005  & 53465  &  ACS/WFC  &  F658N  &  HST~10246/M.~Robberto
\enddata
\end{deluxetable}

\begin{deluxetable}{lcccc}
\tabletypesize{\footnotesize}
\tablewidth{0pt}
\tablecaption{H$\alpha$ Proper motions of knots in HH~517\label{table:propermotions}}
\tablehead{
\colhead{Knot}  &  \colhead{R.A. (J2000)}  &  \colhead{Dec. (J2000)} &
\colhead{V (km s$^{-1}$)} & \colhead{PA} 
}
\startdata					
 s1  & 05:35:18.25  & -05:24:20.9  & 50$\pm$5  & 177  \\
 s2  & 05:35:18.26  & -05:24:21.2  & 55$\pm$5  & 177  \\
 s3  & 05:35:18.30  & -05:24:23.8  & 44$\pm$5  & 170  \\
\enddata
\tablecomments{Coordinates measured on images taken in 2004 (HST Cycle~12).}
\end{deluxetable}

\section{The Orientation of HST-10}

As noted in the introduction, the (projected) major axis of the HST-10 envelope points about 20\arcdeg\ {\em east} of $\theta ^1$Ori~C, and is aligned with the rotation axis of the disk to within a few degrees.  This strongly suggests that the proplyd orientation is dictated by outflows associated with the star/disk system rather than the ionizing radiation from \thetaonec .  Hence, we postulate that the jet and photo-ablation wind from the disk are responsible for the orientation of the HST-10 envelope.  If this is the case, then the combined ram pressure of the jet and disk wind should be larger than the ram pressure of the ionized flow off the IF:  $P_{jet} + P_{wind} > P_{IF}$.
Since ram pressure scales as $n v^2$, the condition can be rewritten as:
\begin{equation}
\frac{n_{jet} v^2_{jet} + n_I v^2_{I}}{ n_{II} v^2_{II}} > 1
\label{eq:ratio}
\end{equation}
where $n$ and $v$ are the density and velocity for the jet, neutral flow (``I''), and IF (``II'').    $v_{I}$ and $v_{II}$ can be approximated by the sound speed in each region:  $v_{I} \approx c_{I} \approx 3$~\kms and  $v_{II} \approx c_{II} \approx 11$~\kms .  From our proper motion analysis of HH~517 we estimate $v_{jet}$ to be $> 50$~\kms , since the actual jet velocity will be higher than the observed proper motion of the shock-front.

The H$\alpha$ surface brightness of the HST-10 IF
can be used to estimate the electron density at the IF using the emission measure and path-length~\citep{Bally:2000}.  Using the  Cycle~12 ACS WFC F658N images, we find a peak value of about 
$2 \times 10^{-12}$~erg~s$^{-1}$~cm$^{-2}$~arcsec$^{-2}$ along the 
northern rim of the proplyd where the HH~517 counter-jet forms a
ring-shaped breakout.  From the observed projected thickness of
the HST-10 IF ($\sim 0.1$\arcsec) the LOS path-length along the
limb-brightened edge of the proplyds is about 200~AU, 
consistent with the apparent thickness of  $\sim 40$~AU for the H$\alpha$
bright layer.  Hence the peak density along the northwestern IF of HST-10 is $n_e \approx 7 \times 10^4$~cm$^{-3}$.  

Since the flow off the disk must also pass (eventually) through the IF, we can calculate $n_I$ from $n_{II}$ using simple mass conservation:  $\dot M_{I} = \dot M_{IF}$  assuming spherical geometry and our adopted velocities.  Assuming spherical geometry, a disk radius of 75~AU, IF distance from the disk of 250~AU, $v_{I} \approx c_{I} \approx 3$~\kms , and  $v_{II} \approx c_{II} \approx 11$~\kms , we estimate a molecular hydrogen density at approximately the disk surface of $n_{I} \approx 2 \times 10^6$~\cmthree .  

The H$\alpha$ surface brightness 
from the HST images can also be used to estimate the electron density in the HH~517 knots.  Using the F658N image, we find $I(H_{\alpha}) = 9 \times 10^{-14}$ erg s$^{-1}$ cm$^{-2}$ arcsec$^{-2}$ and hence $EM = 4.4 \times 10^4$ cm$^{-6}$~pc for HH~517 knot s2.  Assuming spherical symmetry, $L(s2) \approx$ 124 AU, hence $n_e(s2) \approx 9 \times 10^3$~cm$^{-3}$. This can also be considered an {\em upper bound} for the overall jet density, $n_{jet}$, since the HH regions are compressed, post-shock gas.

Inserting the estimated velocities and densities into Equation~\ref{eq:ratio}, we find:
\begin{equation}
\frac{n_{jet} v^2_{jet} + n_I v^2_{I}}{ n_{II} v^2_{II}} \approx 5
\end{equation}
 suggesting that the jet and neutral wind from HST-10 is indeed powerful enough to re-orient the envelope along the disk axis rather than the axis toward \thetaonec. Note, however, that the adopted $n_{jet}$ is an upper bound on the jet density, which drives the ram pressure ratio {\em lower}; and conversely,  $v_{jet}$ is a {\em lower bound} on the jet velocity since it is based on the observed HH object proper motions, which drives the pressure ratio higher (much higher, in fact, since the pressure increases with $v^2$).  Our comparison of the ram pressures shows that a jet and outflow-dominated orientation (as observed) is plausible, though a more detailed analysis and modeling is required to understand how the interaction between the jet and the photo-ablation flow produce the observed IF.  
 
This result suggests that for proplyds far from the ionizing source, envelope orientation may be dominated by ram pressure associated with star/disk outflows, rather than the pressure exerted by the IF.  A review of proplyd orientations in \citet{Robberto:2013} shows, however, that nearly all proplyd orientations are in fact aligned with the ionizing source regardless of projected distance, making HST-10 unique.  HST-10 also has one of the largest silhouette disks among proplyds with bright ionization fronts (the largest silhouette disk in Orion, d114-426, has no obvious IF and is thought to reside in the foreground of the nebula).  If the photo-ablation wind and jet scale up with the disk size, then this suggests that most of the proplyds have disks that are simply too small to re-orient the proplyd envelope and that HST-10 is unique in that it still has a large disk and is close enough to both \thetaonec\ and \thetatwoa\ to produce a bright IF.

\acknowledgements

The data presented herein were obtained at the W.M. Keck Observatory, which is operated as a scientific partnership among the California Institute of Technology, the University of California and the National Aeronautics and Space Administration. The Observatory was made possible by the generous financial support of the W.M. Keck Foundation. We would like to thank Keck Observatory Director Dr. Taft Armandroff for granting a portion of his discretionary time to this project and acknowledge members of Team Keck, with special thanks to Al Conrad, Randy Campbell, and Jim Lyke.
 
The authors wish to recognize and acknowledge the very significant cultural role and reverence that the summit of Mauna Kea has always had within the indigenous Hawaiian community.  We are most fortunate to have the opportunity to conduct observations from this mountain.  

The authors would like to thank the anonymous referee for useful comments which improved the quality of this paper.  



{\it Facilities:} \facility{W. M. Keck Observatory (NIRC2---LGSAO); Hubble Space Telescope (WFPC2, ACS)}

\end{document}